\documentstyle[floats,aps]{revtex}

\newcommand{\forget}[1]{\iffalse#1\fi}
\newcommand{\forgetmenot}[1]{\iftrue#1\fi}

\newcommand{\be}{\begin{equation}}
\newcommand{\ee}{\end{equation}}
\newcommand{\ba}{\begin{eqnarray}}
\newcommand{\ea}{\end{eqnarray}}

\renewcommand{\:}[2]{{\textstyle\frac{#1}{#2}}}
\renewcommand{\;}[2]{{\frac{#1}{#2}}}
\newcommand{\bra}[1]{\left(#1\right)}

\newcommand{\lapp}{\mathrel{\vcenter{\hbox{\tiny \ooalign{\raise 3.25pt
        \hbox{$<$}\crcr $\sim$}}}}}
\newcommand{\gapp}{\mathrel{\vcenter{\hbox{\tiny \ooalign{\raise 3.25pt
        \hbox{$>$}\crcr $\sim$}}}}}
\newcommand{\eqdef}{\!\!\mathrel{\vcenter{\hbox{ \ooalign{\raise 4.75pt
        \hbox{${\textsf{\tiny{def}}}$}\crcr $=$}}}}}

\newcommand{\udot}{\dot u}
\newcommand{\del}{\nabla}
\newcommand{\sdel}{\widetilde\del}

\renewcommand{\>}{\rangle}

\newcommand{\eff}{^{^{_{_{\tiny\mathsf{eff}}}}}}

\title{The Cosmic Microwave Background and Scalar-Tensor Theories of Gravity}
\author{C. A. Clarkson,\footnote{Email:~\texttt{clarkson@mathstat.dal.ca}}
     A. A. Coley\footnote{Email:~\texttt{aac@mathstat.dal.ca}} and
     E. S. D. O'Neill\footnote{Email:~\texttt{oneill@mathstat.dal.ca}}}
\address{Department of Mathematics and Statistics, Dalhousie University, Halifax,
        Nova Scotia, Canada, B3H~3J5.}

\date{\today}

\begin{document}
\maketitle

\begin{abstract}
We show that if all observers see an isotropic cosmic microwave background in
an expanding geodesic perfect fluid spacetime within a scalar-tensor theory of
gravity, then that spacetime must be isotropic and spatially homogeneous. This
result generalizes the Ehlers-Geren-Sachs Theorem of General Relativity, and
serves to underpin the important result that any evolving cosmological model in
a scalar-tensor theory that is compatible with observations must be almost
Friedmann-Lema\^\i tre-Robertson-Walker.
\end{abstract}

\section{Introduction}

The extremely high isotropy of the cosmic microwave background (CMB), which is
isotropic to one part in $10^5$, provides the strongest evidence that the
Universe is isotropic about us. Unfortunately, the spatial homogeneity of the
Universe cannot then be inferred (from our observations of the CMB) without
the Copernican principle (or some equivalent). However, if we assume the
Copernican principle, in that we assume that all observers in the Universe see
the CMB to be as isotropic as we see it, we can then ask: can we infer spatial
homogeneity on the basis of the CMB observations alone? The first attempt to
answer this question within general relativity (GR) resulted in a theorem by
Ehlers, Geren and Sachs (hereafter, EGS;~\cite{egs}) which states that if all
observers in a dust universe see an isotropic radiation field then that
spacetime is spatially homogeneous and isotropic (Friedmann-Lema\^\i
tre-Robertson-Walker, FLRW), in which the isotropic radiation field may be
implicitly identified with the CMB.

The EGS theorem can be generalised trivially to the case of a geodesic and
barotropic perfect fluid \cite{egsother}. However, as has been emphasised
\cite{cla-bar99,fer-95} the resulting spacetime will be FLRW only if the
matter content is of perfect fluid form and the observers are geodesic and
irrotational. Indeed, the EGS theorem has recently been investigated in
inhomogeneous universe models with non-geodesic observers \cite{cla-bar99};
inhomogeneous spacetimes have been found which also allow every observer to
see an isotropic CMB. It has also been shown that a significant subset of
these models are consistent with other observational constraints, regardless
of observer position \cite{bar-cla00}. This means that these models are
consistent with observations even when the Copernican principle is taken into
account~-- and yet the models are significantly inhomogeneous.

The purpose of this paper is to investigate cosmological models which allow an
isotropic radiation field in more general theories of gravity (than GR). In
particular, we shall study scalar-tensor theories of gravity \cite{st,bdt}, in
which gravity is mediated by a long-range scalar field in addition to the usual
tensor fields present in Einstein's theory. Such theories of gravitation, and
especially the simple Brans-Dicke theory of gravity (BDT; \cite{bdt}), are
perhaps the most natural alternatives to GR. Observational limits \cite{wil93}
on the present value of (the Brans-Dicke parameter) $\omega_0$ need not
constrain the value of $\omega$ at early times in more general scalar-tensor
theories (than BDT) with a variable $\omega(\phi)$ \cite{bar-par97}. Hence,
more recently there has been greater focus on the early Universe predictions of
scalar-tensor theories of gravity. Scalar-tensor theories with a `free' scalar
field are perhaps not well motivated since, often, quantum corrections produce
interactions resulting in a non-trivial potential $U(\phi)$.

Scalar-tensor gravity theory is currently of particular interest  since such
theories occur as the low-energy limit in supergravity theories from string
theory \cite{gre-88} and other higher-dimensional gravity theories
\cite{apl-87}. Perhaps cosmology is the ideal setting in which to study
possible stringy effects. However, lacking a full non-perturbative formulation
which allows a description of the early Universe close to the Planck time, it
is necessary to study classical cosmology prior to the GUT epoch by utilizing
the low-energy effective action induced by string theory.  To lowest order in
the inverse string tension the tree-level effective action in four-dimensions
for the massless fields includes the non-minimally coupled graviton, the
scalar dilaton and an antisymmetric rank-two tensor resulting in a
four-dimensional scalar-tensor theory of gravity, thereby generalizing GR by
including other massless fields.  As a result, BDT includes the
dilaton-graviton sector of the string effective action as a special case
$(\omega =-1)$ \cite{gre-88}.

The general form of the extended gravitational action in scalar-tensor
theories is
\be
S = {1\over16\pi} \int d^4x \sqrt{-g} \left[
      \phi R - {\omega(\phi)\over\phi}(\nabla\phi)^2 - U(\phi)
      + 2 {\cal L}_{\rm matter} \right],
\label{eSphi}
\ee
where ${\cal L}_{\rm matter}$ is the Lagrangian for the matter fields, which we
shall assume here corresponds to a comoving perfect fluid. We use units in
which $c= G = 1$. BDT corresponds to the particular choice of $U=0$ and
$\omega=$const. (strictly speaking, $U(\phi)$ must be set to zero in order to
ensure a Newtonian weak field limit). The post-Newtonian parameters of general
relativity are also recovered in the limit that $\omega\to\infty$ and
$(\phi/\omega^3)(d\omega/d\phi)\to0$ \cite{wil93}. The case where $U(\phi)$ is
non-zero but $\omega=0$ is equivalent to higher-order gravity theories
\cite{wan94} with Yukawa type corrections to the Newtonian potential.

The field equations, obtained by varying the action~(\ref{eSphi}) with respect
to the metric and the field $\phi$, are
\ba
G_{ab}&=&\;{8\pi}{\phi}T_{ab}+\;{\omega}{\phi^2}\left(\del_a\phi\del_b\phi-
\:12g_{ab}\del_c\phi\del^c\phi\right)+\phi^{-1}(\del_a\del_b\phi-
g_{ab}\del_c\del^c\phi)-\:12g_{ab}U(\phi).\label{einstein}
\\
(3&+&2\omega) \del_a\del^a\phi  =  8\pi T - \del_a \omega\del^a\phi+
{{dU}\over{d\phi}}
 \label{eEOMphi}
\ea
where the energy-momentum tensor $T^{ab}=2(\delta{\cal L}_{\rm matter}/\delta
g_{ab})$, will take the form of a perfect fluid [see~(\ref{matter}), below]. It
is known that scalar-tensor theories can be rewritten in the conformally
related `Einstein' frame. However, we shall not do this here and we shall work
in the so-called `Jordan' frame. In the scalar-tensor gravity theories the
principle of equivalence is guaranteed by requiring that all matter fields are
minimally coupled to the metric $g_{ab}$. Thus energy-momentum is conserved:
\be
\nabla^{a} T_{ab} = 0.\label{cons}
\ee

In the following section we prove that \emph{in scalar-tensor theories of
gravity, the only geodesic perfect fluid spacetimes admitting an isotropic
radiation field are non-expanding or have a Robertson-Walker (RW) geometry}.
The procedure is to first show that the fluid congruence must be irrotational,
from which it follows from the field equations that the effective energy flux
of the Einstein tensor (discussed below) must vanish. Further study of the
evolution of the effective anisotropic pressure tensor reveals that it too must
vanish or the spacetime must be stationary. Thus we arrive at the above
conclusion.

\section{Spacetimes admitting an isotropic radiation field}

\subsection{Isotropic Radiation and Kinematics}

In GR, assuming that all observers on some congruence $u^a$ see an exactly
isotropic radiation field, then this velocity field has two important
properties:
\ba
\del_{[a}\left(\udot_{b]}-\:13\theta u_{b]}\right)&=&0,\nonumber\\
\sigma_{ab}&=&0,\label{irf}
\ea
where $\theta$, $\udot^a$ and $\sigma_{ab}$ are, respectively, the expansion,
acceleration and the shear  for the timelike congruence $u^a$ (we follow the
notation of \cite{ell-els99} throughout). This `isotropic radiation field
theorem' may be derived from the Einstein-Boltzmann equations for photons in a
curved spacetime \cite{ell-els99,egs,egsother}. As the theorem does not involve
Einstein's field equations, it may be carried over to scalar-tensor theories
without change. The first part of condition~(\ref{irf}) allows us to define
$\del_a Q\equiv\udot_{a}-\:13\theta u_{a}$, for some scalar function $Q$
(proportional to the logarithm of the energy density of the radiation), so we
see that the first condition~(\ref{irf}) is equivalent to
\be
\udot_a=\sdel_a Q,~~~\theta=3\dot Q\label{irf2},
\ee
and it follows that spacetimes admitting an isotropic radiation field are
conformally stationary (as may be seen by considering a conformal
transformation of the congruence, with $e^{Q}$ as the conformal factor
multiplying a stationary metric, with the velocity fields of the two spacetimes
parallel~-- see~\cite{kram}).

We shall restrict our attention to the case in which the physical fluid is
geodesic (i.e., acceleration-free), which then implies that the rotation
of~$u^a$, $\omega_{ab}$, must also vanish: following the argument
in~\cite{cla-bar99}, from~(\ref{irf2}) we can write
\[
\udot_a=\sdel_aQ=0
\]
so that
\be
0=\sdel_{[a}\udot_{b]}=\sdel_{[a}\sdel_{b]}Q= \omega_{ba}\dot Q
=\:13\omega_{ab}\theta,\label{omth=0}
\ee
and we see that $\omega_{ab}=0$ when~$\theta\ne 0$. In this case there exist
comoving coordinates in which the metric takes the form given
in~\cite{col-mac94}.

\subsection{The Matter and the Einstein Tensor}

We are assuming the matter may be described by a perfect fluid, so the energy
momentum tensor has the form:
\be
T_{ab}=\mu u_au_b+\:13\mu h_{ab}+\rho u_au_b+ph_{ab},\label{matter}
\ee
i.e., the matter is a mixture of radiation (which may, without loss of
generality, be considered as a test field, thereby contributing nothing to the
gravitational field) and some other type of perfect fluid satisfying
$\sdel_ap=0$ [which, via~(\ref{cons}), ensures the observers are geodesic]. The
Einstein tensor must then satisfy~(\ref{einstein}), with the matter given
by~(\ref{matter}).

In general, the Einstein tensor can formally be decomposed with respect to a
timelike vector field $u^a$ according to
\be
    G_{ab} \equiv  \mu\eff u_a u_b + p\eff  h_{ab} + q_a\eff u_b
    + q_b\eff u_a + \pi_{ab}\eff=T_{ab}\eff, \label{2.2}
\ee
where
\be
    q_a\eff u^a = 0, \quad h^{ab}\pi\eff_{ab} = 0, \quad \pi_{ab}\eff u^b = 0
    \label{2.3}.
\ee
$T_{ab}\eff$ is now an \emph{effective} energy-momentum tensor, which is not
related to the physical fluid given by~(\ref{matter}) other than
by~(\ref{einstein}). In this formal decomposition $\mu\eff, p\eff, q_a\eff$ and
$\pi_{ab}\eff$ are given by
\ba
    \mu\eff &=&  G_{ab} u^a u^b, \label{2.4}\\
      p\eff &=& \:{1}{3} h^{ab} G_{ab}, \label{2.5}\\
    q_a\eff &=&  -h_a{}^c  G_{cd}  u^d, \label{2.6} \\
    \pi_{ab}\eff &=& h_a{}^c h_b{}^d G_{cd} - \:{1}{3}
                 (h^{cd} G_{cd}) h_{ab}=G_{\<ab\>}. \label{2.7}
\ea
(Angled brackets denote the projected, symmetric, and trace-free part of
tensors, as defined by~(\ref{2.7}); see~\cite{ell-els99}.) For a fluid with
four-velocity $u^a$, these quantities denote, via the field equations, the
energy density, isotropic pressure, energy flux and anisotropic pressure,
respectively, as measured by an observer comoving with the fluid. Using this
decomposition, we may decompose (\ref{einstein}) into the \emph{effective}
energy density and pressure, energy flux and anisotropic pressure. Using this
decomposition of~(\ref{einstein}), given by~(\ref{2.2})~-- (\ref{2.7}), we may
now use the field equations of GR in the 1+3 covariant
formalism~\cite{ell-els99}, but replacing the relevant quantities with their
effective counterparts (i.e., so that $\mu\rightarrow\mu\eff$, etc.).

When $\dot{u}_a$ and~$\omega_{ab}$ are zero,~(\ref{irf}) becomes
\[
\del_{[a}(\theta u_{b]})= u_{[b}\del_{a]}\theta = 0,
\]
which implies (since $\del_a\theta=\sdel_a\theta-\dot{\theta}u_a$) that
\be
\sdel_a\theta=0, \label{gradexp}
\ee
(i.e.,~the expansion is spatially homogeneous). From the constraint equation
relating the divergence of the shear to other kinematical quantities (Eq.,~(32)
in~\cite{ell-els99}) we see that any energy flux component of the Einstein
tensor (or, equivalently, of $T_{ab}\eff$) with respect to the perfect fluid
velocity field must vanish:
\be
q_a\eff=\frac23\sdel_a\theta=0. \label{qa}
\ee
However, from~(\ref{einstein}), and using~(\ref{2.6}), we see that the
effective energy flux of the Einstein tensor is
\be
q_a\eff=-G_{\<a\>b}u^b=\phi^{-1}\bra{\sdel_a\phi\left[\:13\theta-
\omega(\ln\phi)^\cdot\right]
-\sdel_a\dot\phi}=0.\label{q=0}
\ee
Note that we now have a spacetime whose Einstein tensor may be written, using
(\ref{2.4})~-- (\ref{2.7}), as a fluid with zero heat flux with respect to a
geodesic, shearfree and irrotational congruence; hence, the effective
anisotropic pressure vanishing is a necessary and sufficient condition for the
spacetime to be RW~\cite{ell-els99,col-mac94}, and we now show that this must
be the case.

\subsection{Proof of the main result}

From~(\ref{q=0}) we see that these spacetimes must satisfy
\be
\left(\sdel_a\phi\right)^\cdot=
-\omega\left(\ln\phi\right)^\cdot\sdel_a\phi,\label{q=02}
\ee
in order to admit an isotropic radiation field, where we have used the identity
\be
\sdel_a\dot \xi=h_a^{~b}\left(\sdel_b\xi\right)^{\mathbf{\cdot}}-\dot
\xi\udot_a +\:13\theta\sdel_a\xi +\sigma_{a}^{~b}\sdel_b\xi+
\eta_{abc}\omega^b\sdel^c\xi,
\ee
to interchange the order of `time' and `space' derivatives of scalars.
Equation~(\ref{q=02}) may be integrated to give
\be
\sdel_a\phi=\Phi(\phi)F_a,\label{grad_phi}
\ee
where
\be
\Phi(\phi)=\exp\left(-\int\omega\;{d\phi}{\phi}\right),
\ee
and $F^a$ is an arbitrary spatial vector field such that $\dot F^a=0=u^aF_a$.
Now that $\phi$ must satisfy~(\ref{grad_phi}), the total effective anisotropic
pressure tensor
from~(\ref{einstein}) and~(\ref{2.7}), 
\be
\pi_{ab}\eff=\;{1}{\phi}\sdel_{\<a}\sdel_{b\>}\phi
+\;{\omega}{\phi^2}\sdel_{\<a}\phi\sdel_{b\>}\phi,
\ee
becomes
\be
\pi_{ab}\eff=\;{\Phi}{\phi}\sdel_{\<a}F_{b\>}.\label{pi}
\ee

In order to proceed, let us now use the field equations to find further
restrictions on $\phi$. From the evolution equation for the electric Weyl
tensor, and the shear evolution equation, we obtain
\be
E_{ab}=\:12\pi_{ab}\eff
\ee
so therefore
\be
\dot\pi_{\<ab\>}\eff=-\:23\theta\pi_{ab}\eff.\label{pidot}
\ee

However, from~(\ref{pi}) we may calculate the time derivative of $\pi_{ab}$,
\be
\dot\pi_{\<ab\>}\eff=\;{\Phi}{\phi}\left\{-\;{\dot\phi}{\phi}(1+\omega)
\sdel_{\<a}F_{b\>}
+\left(\sdel_{\<a}F_{b\>}\right)^\cdot\right\}\label{pidot2}.
\ee
Using the Ricci identities, and the fact that $R_{a\<bc\>d}u^aF^d=0$ in our
case, we find that
\be
\left(\sdel_{\<a}F_{b\>}\right)^\cdot=-\:13\theta\sdel_{\<a}F_{b\>},
\ee
and so, using~(\ref{pidot},\ref{pidot2}), we find that
\be
\left[\;{\dot\phi}{\phi}(1+\omega)-\;13\theta\right]\sdel_{\<a}F_{b\>}=0.
\ee
Hence we have that $\sdel_{\<a}F_{b\>}=0$, which implies that $\pi_{ab}\eff=0$
and consequently the spacetime is RW, or
\be
\theta=3\;{\dot\phi}{\phi}(1+\omega).\label{theta}
\ee
If we now take the spatial gradient of this expression  we obtain,
using~(\ref{q=0}),
\be
0=\sdel_a\theta=3\omega'\;{\dot\phi}{\phi}\sdel_{a}\phi,
\ee
where $\omega' \equiv \frac{d\omega}{d\phi}$. Now, if $\sdel_a{\phi}=0$, then
$\pi_{ab}\eff=0$, and again the spacetime is RW. If ${\dot\phi}=0$ then the
expansion is zero, and the spacetime is stationary [recall that if $\theta=0$
then the rotation need not vanish~(\ref{omth=0})]. The only other possibility
remaining is that $\omega$ is constant and that $\sdel_a\;{\dot\phi}{\phi}=0$.
Integrating this expression we find that $\phi$ is a separable function (in
time and space in the coordinates of~\cite{col-mac94}), whence on inserting
this into equation (\ref{grad_phi}) we find that for $\omega=$constant$\ne -1$
this leads to $\sdel_a\phi=0$ or $\dot\phi=0$; i.e., the spacetime must again
be either RW or stationary [and in the case $\omega=-1$, (\ref{theta}) ensures
the spacetime is non-expanding].

\subsection{Discussion}

Consequently we have shown that in all cases {\it the spacetime must be
non-expanding or RW}. In particular, all evolving spacetimes must necessarily
have an isotropic and spatially homogeneous geometry. This implies that the
total energy-momentum tensor must be of perfect fluid form whose components
depend on cosmic time only. This does not, however, imply that each of the
separate components of the energy-momentum tensor, such as for example the
scalar field $\phi$, need individually be functions of time
alone~\cite{col-tup}. We note that the preceding calculation did lead to
additional restrictions on $\phi$, namely equations~(\ref{q=02})
and~(\ref{grad_phi}) must be satisfied and $\sdel_{\<a}F_{b\>}=0$. However, if
we take the field equations~(\ref{einstein}) and `contract' them using
$u^{a}u^{b}$ and $g^{ab}$, respectively, using the Brans-Dicke field
equation~(\ref{eEOMphi}) to substitute for $\del_a\del^a\phi$, we obtain two
expressions which can be written, using the equations above, as a sum of terms
each of which has a particular (separable) dependence on time, space or a
functional dependence on $\phi$. Setting these two expressions equal to zero,
it is then straightforward to show that these two equations can only be
satisfied in general for $\phi=\phi(t)$. Hence it follows that all physical
quantities can only depend on time alone.

\section{Conclusions}

We have discussed some of the consequences of the isotropy of the CMB in
universe models in scalar-tensor theories of gravity. We have shown that, if we
assume geodesic motion and a perfect fluid matter source, any expanding
spacetime must have RW geometry, which is also the situation in GR. As is the
case in GR, we would expect that this result will have a perturbed equivalent,
in that the near isotropy of the CMB will imply a nearly homogeneous and
isotropic cosmological model in these alternative theories of
gravity~\cite{almost}. Thus we conclude that if we wish to consider
cosmological implications of scalar tensor theories of gravity, CMB
observations imply that the canonical `almost' FLRW models will be the correct
models to use, provided that other observations fulfill the assumptions used
here.

\acknowledgments

We would like to thank Roy Maartens for useful comments. This work was funded
in part  by NSERC of Canada.

\end{document}